\documentclass[reprint,aps,amsmath,amssymb,notitlepage,superscriptaddress, twocolumn,color,epsfig,graphicx,bm,floatfix,nofootinbib,longbibliography]{revtex4-1}
\pdfoutput=1
\usepackage{braket}
\usepackage[utf8]{inputenc}
\usepackage[english]{babel}
\usepackage[T1]{fontenc}
\usepackage{amsmath}
\usepackage{hyperref}
\usepackage{multirow}
\usepackage{tikz}
\usepackage{lipsum}
\usepackage{enumitem}

\usepackage[utf8]{inputenc}

\usepackage{tikz}

\usetikzlibrary{decorations.markings}

\usepackage{amsmath}  
\usepackage{url}
\usepackage{eqnarray}
\usepackage{hyperref}
\usepackage{cleveref}
\usepackage{mathtools}
\usepackage{siunitx}
\usepackage{multirow}
\usepackage{microtype}
\usepackage{array}
\usepackage{tabulary}
\usepackage{longtable}
\usepackage[version=4]{mhchem}
\usepackage{graphicx}
\newcolumntype{K}[1]{>{\centering\arraybackslash}p{#1}}

\usepackage{graphicx}   
\usepackage{verbatim}   
\usepackage{color}      
\usepackage{subfigure}  
\usepackage{hyperref}   

\usepackage{lipsum}

\begin{document}

\title{Constrained Quantum Optimization at Utility Scale:\\ Application to the Knapsack Problem}

\author {Naeimeh Mohseni}
\email{naeimeh.mohseni@eon.com}
\affiliation {E.ON Digital Technology GmbH, Hannover, Germany}
\author{Julien-Pierre Houle}\thanks{Equal contribution.}
\affiliation{University of Calgary, Quantum City, Calgary, AB T2N 1N4, Canada}
\affiliation{Université de Sherbrooke, Institut Quantique, Sherbrooke, QC J1K 2R1, Canada}
\author {Ibrahim Shehzad}\thanks{Equal contribution.}
\affiliation{IBM Quantum, Thomas J. Watson Research Center, Yorktown Heights, NY 10598, USA}

\author {Giorgio Cortiana}
\author {Corey O'Meara}
\affiliation {E.ON Digital Technology GmbH, Hannover, Germany}
\author{Adam Bene Watts}
\email{adam.benewatts@ucalgary.ca}
\affiliation{University of Calgary, Quantum City, Calgary, AB T2N 1N4, Canada}
\begin{abstract}
Constrained combinatorial optimization problems are challenging for quantum computing, particularly at utility-relevant scales and on near-term hardware. At the same time, these problems are of practical significance in industry; for example, the Unit Commitment (UC) problem in energy systems involves complex operational constraints. To address this challenge, we apply copula-QAOA (cop-QAOA), a hardware-efficient approach for constrained optimization to a single-period UC that can be reduced to a one-dimensional knapsack.   Cop-QAOA biases the quantum state toward feasible solutions using constant-depth mixers and appropriately biased initial states. We implement our benchmark on problem instances that are confirmed to be hard for classical solvers such as Gurobi. Our results show that cop-QAOA often finds solutions better than a lazy greedy baseline and very close to — and in some instances surpasses — those obtained by Gurobi, with only a few QAOA rounds. 
This work presents the largest successful demonstration of the knapsack problem on IBM Quantum hardware using up to 150 qubits, and more generally, the largest demonstration of constrained combinatorial optimization where constraints are enforced via shallow mixers.
\end{abstract}
\maketitle
\section{Introduction}
Unit Commitment (UC) \cite{agliardi2024machine} sits at the heart of short-term power system operations. The task is to decide which generating units should be online, and when, so that the system can reliably supply the predicted demand while honoring engineering and operational limits \cite{mohseni2025evidence}. As modern grids incorporate fluctuating renewable sources and increasingly complex market structures, the UC problem becomes both larger and more intricate. Improving the efficiency and robustness of UC solvers is therefore essential for grid operators, since even small advances can reduce operational costs and enhance system stability. This broader context motivates exploring new algorithmic techniques,  that may offer advantages for tightly constrained scheduling problems. 

UC and related power system optimization problems have also been studied using quantum approaches, including variational quantum algorithms \cite{koretsky2021adapting} and methods based on quantum kernels \cite{agliardi2025mitigating, agliardi2024machine}. In this work, we focus on a simplified, single-period version of the UC problem obtained by omitting all time-coupling constraints such as ramping limits and minimum up/down times \cite{koretsky2021adapting}. While simplified, this formulation already captures the core combinatorial structure of UC.

Variational quantum algorithms, such as the Quantum Approximate Optimization Algorithm (QAOA), are natural candidates for tackling combinatorial optimization problems. However, their direct application to the UC is limited by several factors. First, the UC problem is a mixed binary optimization problem, requiring optimization over both binary and continuous variables. There are large time and space overheads associated with manipulating continuous variables on quantum hardware, making it challenging to optimize over these variables using QAOA-like techniques on near-term quantum devices. Second, the UC problem is a constrained optimization problem, meaning the feasible solution space for the problem occupies only a fraction of the full Hilbert space typically explored by variational algorithms. In this work we address both of these limitations. 

The first limitation has been observed in prior work on quantum approaches to the UC problem. In particular,~\cite{koretsky2021adapting} observed this limitation and proposed a hybrid quantum-classical workflow, with a classical outer loop optimizing over continuous variables and a quantum inner loop solving a purely binary variable assignment problem. This binary variable assignment problem is equivalent to the (1D) knapsack problem; thus, this hybrid workflow has the potential to outperform purely classical approaches provided quantum computers can offer some advantage in solving knapsack problems. One downside of the approach in~\cite{koretsky2021adapting} is that it requires the classical outer loop to optimize over one additional continuous variable for every new unit in the original UC problem, which becomes infeasible at large problem sizes. In this work, we modify this approach and show the (single-timestep) UC problem can be solved using a hybrid workflow with only one continuous parameter. The main innovation is the use of first order optimality conditions to reduce the number of continuous variables in the problem. This approach is discussed in more detail in~\Cref{subsec:reducing UC to knapsack}, with formal proofs provided in~\Cref{app:derivative}.

A number of techniques have also been proposed to modify QAOA for constrained optimization problems. Several extensions address constraints by modifying the mixing Hamiltonian to limit the search through feasible space \cite{hadfield2019quantum, wang2020xy, bartschi2020grover}. While effective in principle, these approaches often require increased circuit depth or more complex initial state preparation, which can be prohibitive on near-term quantum hardware. An alternative strategy is to bias the quantum state toward the feasible subspace without strictly enforcing feasibility, using biased initial states and constant-depth mixers \cite{van2021quantum}. This approach encourages exploration of feasible or near-feasible configurations while maintaining a hardware-efficient circuit structure, offering a practical compromise between constraint handling and circuit complexity on near-term quantum devices. This is the approach we follow in this paper. To our knowledge, this paper gives the first utility-scale investigation of this strategy on real quantum hardware. 





A central goal of this work is to push constrained quantum optimization to utility-scale problem sizes on real hardware and to assess its performance in the presence of noise. We show that cop-QAOA \cite{van2021quantum} enables successful implementation of constrained optimization problems on IBM quantum hardware, producing solutions that are competitive with those obtained using classical solvers such as Gurobi. At the same time, we encounter challenges in training the variational parameters, highlighting the need for additional heuristics tailored to constrained optimization, or for extending existing techniques—such as CVaR-based methods \cite{Barkoutsos2020improving} — to settings in which constraints are enforced structurally through biased initial states and finite-depth mixers.  

As a final remark, beyond algorithmic design, benchmarking quantum optimization methods requires careful consideration of the problem instances themselves. Relying solely on worst-case complexity arguments or on instance classes labeled as “hard” does not guarantee practical hardness for state-of-the-art classical solvers. Indeed, although the single-period UC problem is NP-hard \cite{bendotti2019complexity}, we find that generating instances that remain challenging for solvers such as Gurobi (at scales below 150 units) is nontrivial. This underscores the importance of instance-level analysis when evaluating potential quantum advantages particularly at such scale. Here, we benchmark on carefully selected instances  that are particularly challenging for Gurobi, such that it cannot find a solution with zero optimality gap. For these instances, we show that cop-QAOA can successfully compete with Gurobi and even identify a solution that is slightly better.

\section{Problem Formulation}

A simplified, single-period version of the UC problem can be written by omitting all time-coupling constraints such as ramping and minimum up/down times \cite{koretsky2021adapting}. In this reduced setting, each generating unit $i$ is characterized by a fixed commitment cost, a linear production cost, and a quadratic production cost. 
\begin{equation}
    C(\vec{y}, \vec{p}):=\sum_{i=1}^n\left(A_i \cdot y_i+B_i \cdot p_i+C_i \cdot p_i^2\right),
    \label{eq:UC_cost}
\end{equation}
subject to 
\begin{equation}
\label{eq:constraints}
\begin{aligned}
\sum_{i=1}^n p_i &\geq L, \quad \forall t \quad \text { (demand constraint) }, \\
p_{i, \min } \cdot y_i & \leq p_i \leq p_{i, \max } \cdot y_i, \quad \forall i \quad \text { (generation limits), } \\
y_i & \in\{0,1\} \\
p_i & \in \mathcal{R} .
\end{aligned}
\end{equation}

where $A_i$, $B_i$,  and $C_i \in \mathbb{R}$ are constants related to the properties of the power system.
In this formulation, $y_i$ determines whether unit $i$ is committed (on) or not, while $p_i$  represents the power output of each unit which is turned on. The output bounds ensure that a unit $i$ can only output power if it is turned on and, it outputs power in the range $[p_{i, min},p_{i, max}]$.

\subsection{Reducing single-period UC to 1D-Knapsack}
\label{subsec:reducing UC to knapsack}

In this section we show how single-period UC can be converted to optimization problem involving instances of the 1D knapsack problem and a single additional continuous parameter. 
The  1D knapsack problem considers $n$ items with integer weights $w_1, \ldots, w_n$ and values $v_1, \ldots, v_n$, together with a capacity $c \in \mathbb{N}$. The task is to choose a subset of items whose total value is maximized without exceeding the capacity constraint. The problem is expressed as finding a binary vector $x^*=\left(x_1, \ldots, x_n\right) \in\{0,1\}^n$ that solves

\begin{equation}
\label{eq:optimize}
\max \left(\sum_{i=1}^n x_i v_i\right)
\end{equation}

subject to

\begin{equation}
    \sum_{i=1}^n x_i w_i \leq c, \quad x_i \in\{0,1\}.
    \label{eq:constraint_KP}
\end{equation}

We start from the observation that a solution $(\vec{y}, \vec{p})$ minimizing the UC cost function given in \Cref{eq:UC_cost} must satisfy
\begin{align}
    \left(\frac{\partial C}{\partial p_i}\right)\bigg|_{p_i} = \left(\frac{\partial C}{\partial p_j}\right)\bigg|_{p_j} \label{eq:first_order_optimality}
\end{align}
for any $i,j$ with $p_i \notin \{y_i p_{i,\min},y_i p_{i,\max}\}$ and $p_j \notin \{y_j p_{j,\min},y_j p_{j,\max}\}$. This follows directly from calculus, and can be understood intuitively by noting that, if this condition is not satisfied, a lower-cost solution outputting the same power can be achieved by infinitesimally lowering $p_i$ and raising $p_j$, or vice-versa. 

Following the same logic, and observing that $(\partial C/ \partial p_i)$ is a monotonically increasing function of $p_i$ for all $i$ implies that there exists a real number $\mathcal{D}$ such that one of the equations
\begin{enumerate}[label = (\roman*)]
    \item $\left(\frac{\partial C}{\partial p_i}\right)\bigg|_{p_i} = \mathcal{D}$ \label{itm:derivative_cond_1}
    \item $\left(\frac{\partial C}{\partial p_i}\right)\bigg|_{p_i} > \mathcal{D}$ and $p_i = p_{i,min}$ \label{itm:derivative_cond_2}
    \item $\left(\frac{\partial C}{\partial p_i}\right)\bigg|_{p_i} < \mathcal{D}$ and $p_i = p_{i,max}$ \label{itm:derivative_cond_3}
\end{enumerate}
is satisfied for all $p_i$ with $y_i \neq 0$. A formal derivation of this fact is given in Appendix A.

We can use the conditions above to analytically eliminate most of the continuous variables $p_i$ appearing in the single timestep UC problem. 
For any fixed marginal-cost parameter $\mathcal{D}$, the stationarity condition
\begin{equation}
\left(\frac{\partial C}{\partial p_i}\right)=B_i+2 C_i p_i=\mathcal{D}
\end{equation}
gives the interior solution
\begin{equation}
p_i=\frac{\mathcal{D}-B_i}{2 C_i}
\end{equation}
which can then be projected onto the feasible interval $\left[p_{i, \text { min }}, p_{i, \text { max }}\right]$.
Once the $p_i$ are determined, the problem reduces to choosing the subset of units that minimizes

\begin{equation}
\sum_{i=1}^n\left(A_i+B_i p_i+C_i p_i^2\right) y_i \quad \text { subject to } \quad \sum_{i=1}^n p_i y_i \geq L,
\label{eq:UC_problem}
\end{equation}
a purely binary optimization problem parameterized by $\mathcal{D}$. After rewriting in terms of variables $z_i = 1-p_i$, this problem is equivalent to finding
    \begin{align}
        \arg \max \left(-\sum_i A_i z_i + B_i p_i z_i + C_i p_i^2 z_i\right) \\
        \text{s.t.}
        \sum p_i z_i \leq L - \sum_i p_i.
        \label{eq:UC_inverse_mapping}
    \end{align}
This is exactly an instance of the knapsack problem, with weights $w_i = p_i$ and values $v_i = -(A_i + B_ip_i + C_ip_i^2)$. 

Solving instances of the knapsack problem for each possible value of $\mathcal{D}$ and searching over $\mathcal{D}$  identifies the value for which the demand constraint is met and the marginal-cost balance holds.

\subsection{Numerical Investigation of the Knapsack Reduction}

\begin{figure}
    \centering
    \includegraphics[width=\linewidth]{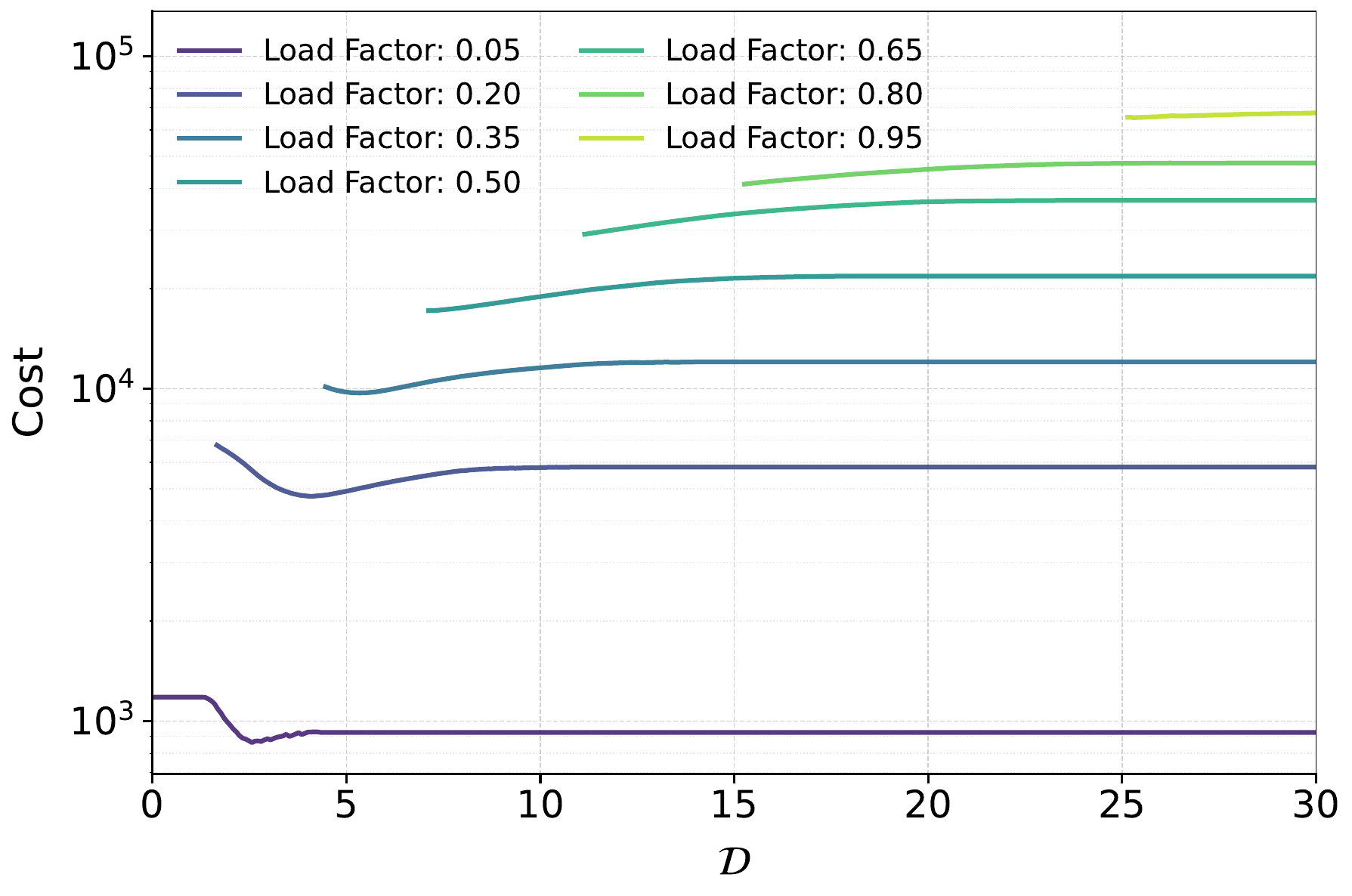}
    \caption{Impact of the parameter $\mathcal{D}$ on the quality of the solution find when mapping the UC problem to a Knapsack like problem. Figure shows the UC cost function (eq. \ref{eq:UC_cost}) as a value of $\mathcal{D}$ when solving problems with different values of load factor. The load factor is the ratio $L / \sum_i p_{max, i}$. Problem instances are generated for 100 units with randomized parameters drawn from the following ranges: $A \in [10, 50]$, $B \in [0.5, 1.5]$, $C \in [0.01, 0.2]$, $p_{\min} \in [10, 20]$, and $p_{\max} \in [50, 100]$.}
    \label{fig:fig1}
\end{figure}

In this section we investigate practical aspects of the reduction presented in the previous section.
In \Cref{fig:fig1}, we show numerical observations regarding the optimization landscape of $\mathcal{D}$. We observe that the cost function exhibits a convex shape with respect to $\mathcal{D}$ across all load factors. This convexity suggests that optimization over $\mathcal{D}$ may be straightforward, with methods such as bisection converging efficiently.


\subsection{Hardness of the Single-Timestep UC Problem}

Mathematically, the single-timestep UC problem is at least as hard as the 1D knapsack problem. In particular, taking an instance of the single timestep UC problem and setting variables $p_{i,\min} = p_{i,\max}$ for all $i$ immediately recovers an instance of the 1D knapsack problem. Thus, the single timestep UC problem is NP-hard, and exactly solving all instances of that should be infeasible on both quantum and classical computers. 

In practice, however, most natural instances of the single timestep UC problem are easy to solve classically. In our numerical experiments, we found that Gurobi could quickly solve instances of the single-timestep UC problem with up to 400 units. This work does not aim to find a practical quantum advantage for solving single-timestep UC problems. Instead, a central goal of this work is to identify useful techniques which may generalize to other, more involved versions of the UC problem. 

For the reasons above, in the later half of the paper we focus on techniques for solving instances the 1D knapsack problem on a quantum computer, without discussing how the single-timestep UC problem might reduce to those instances.

\section{Classical Knapsack Solvers \label{sec:classical solvers}}
Here we describe briefly the classical algorithms we used to solve our knapsack instances. \\
\textbf{Lazy greedy}: For each item $i$, the ratio is defined as $r_i=v_i / w_i$. The lazy greedy method \cite{van2021quantum} orders all items in non-increasing ratio and then selects them sequentially until the capacity constraint is about to be violated. Once the next item in the ordered list would exceed the capacity, the algorithm terminates without considering less efficient items that might still fit. When expressed as a binary decision vector ordered by decreasing ratio, the resulting solution consists of a prefix of selected items followed by unselected ones, i.e., a contiguous block of ones followed by zeros. As it is suggested in \cite{van2021quantum}, we construct a warm-start distribution from the lazy greedy heuristic, which induces a sharp threshold when items are ordered by their efficiency ratios $r_i$. To smoothen this step-like behavior, we assign each item a selection probability

\begin{equation}
 p_i=\frac{1}{1+C e^{-k\left(r_i-r^*\right)}}  \label{EQ:proba} 
\end{equation}

where the parameters $k, C$, and $r^*$ control the steepness and location of the transition. In the limit $k \rightarrow \infty$ with $r^*$ set to the lazy greedy stopping ratio, this construction recovers the original lazy greedy solution. We set $C=\sum_i w_i / c-1$ where $c$ is the capacity of the knapsack.\\
\textbf{Gurobi}:
We also use Gurobi \cite{gurobi} a commercial optimization solver widely used for linear, mixed-integer, and quadratic programs. This solver employs a branch-and-cut framework, which integrates branch-and-bound with cutting-plane techniques to progressively tighten relaxations while systematically exploring the solution space. During optimization, Gurobi reports an optimality gap defined as the difference between the best feasible solution found and the strongest upper bound obtained from the corresponding dual relaxation. An optimality gap of zero certifies global optimality.
\section{Copula QAOA }
The Quantum Approximate Optimization Algorithm (QAOA) \cite{farhi2014qaoa} is a variational quantum algorithm that alternates between applying a cost Hamiltonian and a mixer Hamiltonian to a parameterized quantum state.  QAOA faces a fundamental limitation on constrained problems such as knapsack: the feasible solution space occupies only a  fraction of the full Hilbert space, and the standard transverse-field mixer freely explores both valid and invalid configurations. 
QAOA has been extended to handle constrained optimization problems through the generalization of the mixing Hamiltonian, leading to the so-called Quantum Alternating Operator Ansatz (QAOA) algorithms \cite{hadfield2019quantum}.  Variants of the mixing Hamiltonian are designed to preserve the feasible subspace for specific constraint problems \cite{hadfield2019quantum, wang2020xy, bartschi2020grover}. However, some of them often increase the complexity in terms of mixing depth. 
One such approach is copula-QAOA (cop-QAOA) \cite{van2021quantum}, which, like the original QAOA, uses a constant-depth mixer   and initial state preparation circuit. Cop-QAOA biases the final quantum state towards feasible solutions for problems with linear constraints, without strictly restricting the state to the feasible subspace—a task that typically requires complex circuits. Instead, the algorithm keeps the state near the feasible subspace while maintaining implementable circuit depth. The initial states and the mixing Hamiltonian are biased as follows

\textbf{Biased initial state}
In standard QAOA, the initial state is a uniform superposition over all $2^n$ configurations:

$$
\frac{1}{\sqrt{2^n}} \sum_{x \in\{0,1\}^n}|x\rangle .
$$

In cop-QAOA, the initial state is biased toward promising solutions by assigning each qubit $i$ a probability $p_i$ of being in the $|1\rangle$ state:

\begin{equation}
\left\{\begin{array}{l}
\left|p_i\right\rangle:=\sqrt{1-p_i}|0\rangle+\sqrt{p_i}|1\rangle, \\
\left|p_i^{\perp}\right\rangle:=-\sqrt{p_i}|0\rangle+\sqrt{1-p_i}|1\rangle .
\end{array}\right.
\end{equation}

The vector $\left|p_i\right\rangle$ is used to construct the initial product state across all qubits. The probabilities $p_i$ can be set based on a warm-start solution, prior knowledge, or heuristics to bias the system toward feasible configurations. We use the smoothed lazy greedy heuristic described in Sec.~\ref{sec:classical solvers} as our warm-start strategy. Therefore, the initial state is,
\begin{equation}
|\psi_{in}\rangle=\bigotimes_{i=1}^n\left|p_i\right\rangle
\end{equation}

where $p_i$ is defined in Eq. \ref{EQ:proba}.

\textbf{Biased mixing Hamiltonian}
To further guide the state toward feasible solutions while allowing correlations between variables, reference \cite{van2021quantum} introduces a family of copula mixers. For a pair of qubits with marginal probabilities $p_1$ and $p_2$ and a correlation parameter $\theta \in[-1,1]$, the copula distribution over the two bits is defined as \cite{van2021quantum}:

$$
\begin{aligned}
& p_{\text {cop }}(0,0)=\left(1-p_1\right)\left(1-p_2\right)+\theta p_1 p_2\left(1-p_1\right)\left(1-p_2\right), \\
& p_{\text {cop }}(0,1)=\left(1-p_1\right) p_2-\theta p_1 p_2\left(1-p_1\right)\left(1-p_2\right), \\
& p_{\text {cop }}(1,0)=p_1\left(1-p_2\right)-\theta p_1 p_2\left(1-p_1\right)\left(1-p_2\right), \\
& p_{\text {cop }}(1,1)=p_1 p_2+\theta p_1 p_2\left(1-p_1\right)\left(1-p_2\right) .
\end{aligned}
$$

The parameter $\theta$ controls the strength and type of correlation induced between the two qubits: $\theta>0$ introduces positive correlation, $\theta<0$ introduces anti-correlation, and $\theta=0$ reduces to independent sampling according to the marginals. Following the findings in \cite{van2021quantum}, which demonstrated that $\theta=-1$ is consistently optimal across problem instances, we fix $\theta = -1$ in our implementation to maximize the anti-correlations. 


To implement this correlated distribution at the circuit level, one constructs a two-qubit unitary $\mathbf{R}\left(p_i, p_{i^{\prime}}\right)$ such that

$$
\mathbf{R}_{\text{Cop}}\left(p_i, p_{i^{\prime}}\right)|00\rangle=\sum_{x_i, x_{i\prime} \in\{0,1\}} \sqrt{p_{\text {cop }}\left(x_i, x_{i^{\prime}}\right)}\left|x_i x_{i^{\prime}}\right\rangle,
$$

Explicitly, $\mathbf{R}$ can be implemented as
\begin{equation}
\begin{aligned}
\mathbf{R}_{\mathrm{Cop}}\left(p_i, p_{i^{\prime}}\right)= & \bar{C}_i\left(\mathrm{RY}_{i^{\prime}}\left(2 \sin ^{-1} \sqrt{p_{i^{\prime} \mid \bar{i}}}\right)\right) \\
& \times C_i\left(\mathrm{RY}_{i^{\prime}}\left(2 \sin ^{-1} \sqrt{p_{i^{\prime} \mid i}}\right)\right) \\
& \times \mathrm{RY}_i\left(2 \sin ^{-1} \sqrt{p_i}\right),
\end{aligned}
\end{equation}
with 
\begin{equation}
\begin{aligned}
& p_{i^{\prime} \mid i}=p_{i^{\prime}}\left(1-\left(1-p_i\right)\left(1-p_{i^{\prime}}\right)\right), \\
& p_{i^{\prime} \mid \bar{i}}=p_{i^{\prime}}\left(1+p_i\left(1-p_{i^{\prime}}\right)\right),
\end{aligned}
\end{equation}
where $\mathrm{RY}_i(\alpha)=\exp(-i \alpha Y_i/2)$ is a rotation about the Pauli-$Y$ axis on qubit $i$, and $C_i$, $\bar{C}_i$ are controlled operations on qubit $i$ conditioned on $\ket{1}$ and $\ket{0}$, respectively.
and then we can define the mixer Hamiltonian as
\begin{equation}
H_{\text {cop }}=R_{\text {cop }} (p_i,p_{i'} )(Z_i +Z_i') R^{\dagger}_{\text {cop }} (p_i,p_{i'} )
\end{equation}
where $Z_i$  is the Pauli-Z operator acting on qubit $i$. This Hamiltonian generates the copula mixer unitary for each layer $l$ as

\begin{equation}
U_{\mathrm{M}}^{\text{Cop}}(\beta_l)=\exp \left(-i \beta_l H_{\text {cop }}\right)
\end{equation}

which governs transitions within the feasible region of the solution space.

\textbf{Cost Hamiltonian} The optimization objective  for our knapsack problem is encoded through a diagonal cost Hamiltonian $H_C=\sum_i v_i Z_i$, which assigns a phase proportional to the cost of each computational basis state. The associated cost unitary in layer $l$ is

\begin{equation}
U_C(\gamma_l)=\exp \left(-i \gamma_l H_c\right)
\end{equation}
After $p$ alternating applications of the cost and copula mixer unitaries, the resulting variational state is 
\begin{equation}
|\psi(\boldsymbol{\beta}, \boldsymbol{\gamma})\rangle=\prod_{\ell=1}^p U_{\mathrm{M}}^{\mathrm{Cop}}\left(\beta_{\ell}\right) U_C\left(\gamma_{\ell}\right)\left|\psi_{\mathrm{init}}\right\rangle .
\end{equation}
The parameters $\beta$ and $\gamma$ are chosen to minimize the expectation value of the cost Hamiltonian, $ C(\boldsymbol{\beta}, \boldsymbol{\gamma})=\langle\psi(\boldsymbol{\beta}, \boldsymbol{\gamma})| H_C|\psi(\boldsymbol{\beta}, \boldsymbol{\gamma})\rangle$, which serves as the classical objective function in the variational optimization.

\section{Previous Relevant Works}
Cop-QAOA has previously been applied to the knapsack problem in two studies \cite{van2021quantum, christiansen2024quantum}, both of which focus on generating hard knapsack  instances using methods introduced in \cite{pisinger2005hard, jooken2022new}. Both
studies are constrained to small problem sizes, typically
fewer than 30 items. The reference \cite{christiansen2024quantum} reports that even for these small size instances, the performance of cop-QAOA does not scale favorably with increasing numbers of items or QAOA rounds. 

Here, we consider the same instance-generation methods introduced in \cite{pisinger2005hard, jooken2022new}. We  demonstrate a successful implementation of cop-QAOA for instances with up to 150 items on both simulator and quantum hardware. We conjecture that the observed  limitation in \cite{christiansen2024quantum} is related to the difficulty of identifying optimal variational parameters. Consistent with this, we also identified instances  for which finding optimal variational was very challenging. We discuss this more in detail in Sec. \ref{training}.
\section{Technical Remarks}
\subsection{Problem Instance Selection \label{subsec:instance}}
Following systematic construction recipes proposed in previous studies \cite{pisinger2005hard, jooken2022new}, we created several families of instances designed to reproduce the reported hardness and evaluated their hardness for Gurobi.

In practice, we found that Gurobi could certify optimal solutions with zero optimality gap for nearly all instances based on \cite{pisinger2005hard}, and that the runtime did not scale badly with item size (up to 150 items). Only a small subset of instances generated according to \cite{jooken2022new} posed difficulties for optimality verification within this size range.

Our benchmarks here therefore focus on two families of instances. The first family consists of instances drawn from the Inversely Strongly Correlated distribution, following the construction described in \cite{pisinger2005hard}.  In this distribution, item values $v_j$ are sampled uniformly from $\{1, \ldots, 1000\}$, and the corresponding weights are drawn from the interval $\left\{v_j+98, \ldots, v_j+102\right\}$. These instances are known to be challenging for classical solvers due to the strong correlation between weights and values, which leads to poor conditioning and reduces the effectiveness of greedy heuristics and branch-and-bound methods. The knapsack capacity is set as $c \leftarrow\left\lceil\alpha \cdot\left(\sum_i w_i\right) / 100\right\rceil$ where $\alpha$ is a uniformly random value in $[10,11,...,20]$. Nevertheless, for instances of this type with up to 150 items, Gurobi was almost always able to certify optimal solutions with zero optimality gap indicating that these instances are not particularly challenging for state-of-the-art classical solvers at this scale.

The second family consists of instances generated following the construction of \cite{jooken2022new}. We generated a number  of such instances and empirically identified a small subset that is particularly challenging for Gurobi at this scale (smaller than 150 items). Details of the instance generation procedure and the corresponding implementation can be found in \cite{jooken2022new}. Interpreting the construction as a unit commitment knapsack mapping, the method can be viewed as producing weights contained within discretized generation blocks in one fixed unit system (e.g. megawatts) together with costs of comparable order within each block-size layer, plus smaller heterogeneous items corresponding to fine dispatch adjustments.



\subsection{Benchmarking Metrics}
 As metrics for comparison, we consider the best-found solution, the approximation ratio, and the ratio of valid solutions.  When the number of feasible solutions returned by cop-QAOA at different depths and the lazy greedy differ substantially, we compute the approximation ratio using the top $k$ samples.

 We define the approximation ratio as

\begin{equation}
A r=\frac{\sum_{\text {feasible } b}\operatorname{cost}(b) \operatorname{count}(b)}{N_{\text {feasible }} C^{\star}} .
\end{equation}

  where $b$ is a feasible bitstring, $\operatorname{cost}(b)$ is its cost value, $\operatorname{count}(b)$ is how many times it appeared in the measurement outcomes, $C^{*}$ is the optimal cost value. $N_{\text {feasible }}$ denotes the number of feasible measurement outcomes (shots), i.e., $N_{\text {feasible }}= \sum_{b \in F} \operatorname{count}(b)$.

\subsection{Training Variational Parameters \label{training}}
We begin with a depth $p=1$ QAOA circuit and initialize the parameters $\beta$ and $\gamma$ randomly. We run the QAOA optimization 20 times and evaluate the cost function $C(\boldsymbol{\beta}, \boldsymbol{\gamma})$ over sampled bit strings for each run. The parameters that minimize this average are taken as the optimal values for the first layer. Once these optimal parameters have been identified, we fix them and move to depth $p=2$. At this stage, the first layer's $\beta$ and $\gamma$ remain fixed, and we train only the new variational parameters introduced for the second layer. In general, at each depth $p$, only the newly added $\beta_p$ and $\gamma_p$ are trained, while all parameters from earlier layers are kept fixed.

In practice, the objective function $C(\beta, \gamma)$ is evaluated by sampling bit strings from the variational state $|\psi(\boldsymbol{\beta}, \boldsymbol{\gamma})\rangle$ and computing their classical objective values. Suggested by  \cite{van2021quantum}, each sampled bit string is first tested for feasibility, and configurations that violate the constraints are assigned an objective value of zero. The average over the cost value of the all valid sampled values provides an estimate of $C(\beta, \gamma)$, which is then minimized by a classical optimizer.

For certain problem instances, we observed that starting from a random initialization of $\gamma$ and $\beta$ and optimizing the average cost function does not yield parameters that allow cop-QAOA to surpass the lazy greedy baseline. This behavior is expected, as improvements in the expectation value of the cost Hamiltonian do not necessarily translate into improvements in the best sampled solution. In some instances, to obtain solutions that outperform the lazy greedy baseline, we therefore performed a two dimensional grid search over the best-observed cost as a function of $\gamma$ and $\beta$ at depth $p=1$, and used the parameters corresponding to the best value to initialize the first layer. Training from this initialization then leads to improved solutions.

Here we  assign zero value to cost to infeasible configurations which can induce extended flat regions in the effective objective function, as many sampled configurations yield identical objective values. Such plateaus reduce sensitivity of the estimated objective to changes in the variational parameters, making optimization more difficult. One possible alternative, instead of assigning zero cost to infeasible solutions, is to post-process invalid bit strings by mapping them to feasible configurations.  In such setting, tail-focused objectives such as CVaR, which rely on  variation in the upper tail of the objective distribution, are not guaranteed to improve performance. This suggests that more advanced optimization strategies that explicitly incorporate constraint satisfaction may be required for constrained optimization. 

To run on our experiments on hardware, we do not re-train cop-QAOA. Instead, we use the parameters optimized classically on the simulator and then execute the circuit with these fixed parameters on the physical device, as in \cite{barron2024provable}.


\begin{figure*}[t]
    \centering
    \includegraphics[width=0.98\linewidth]{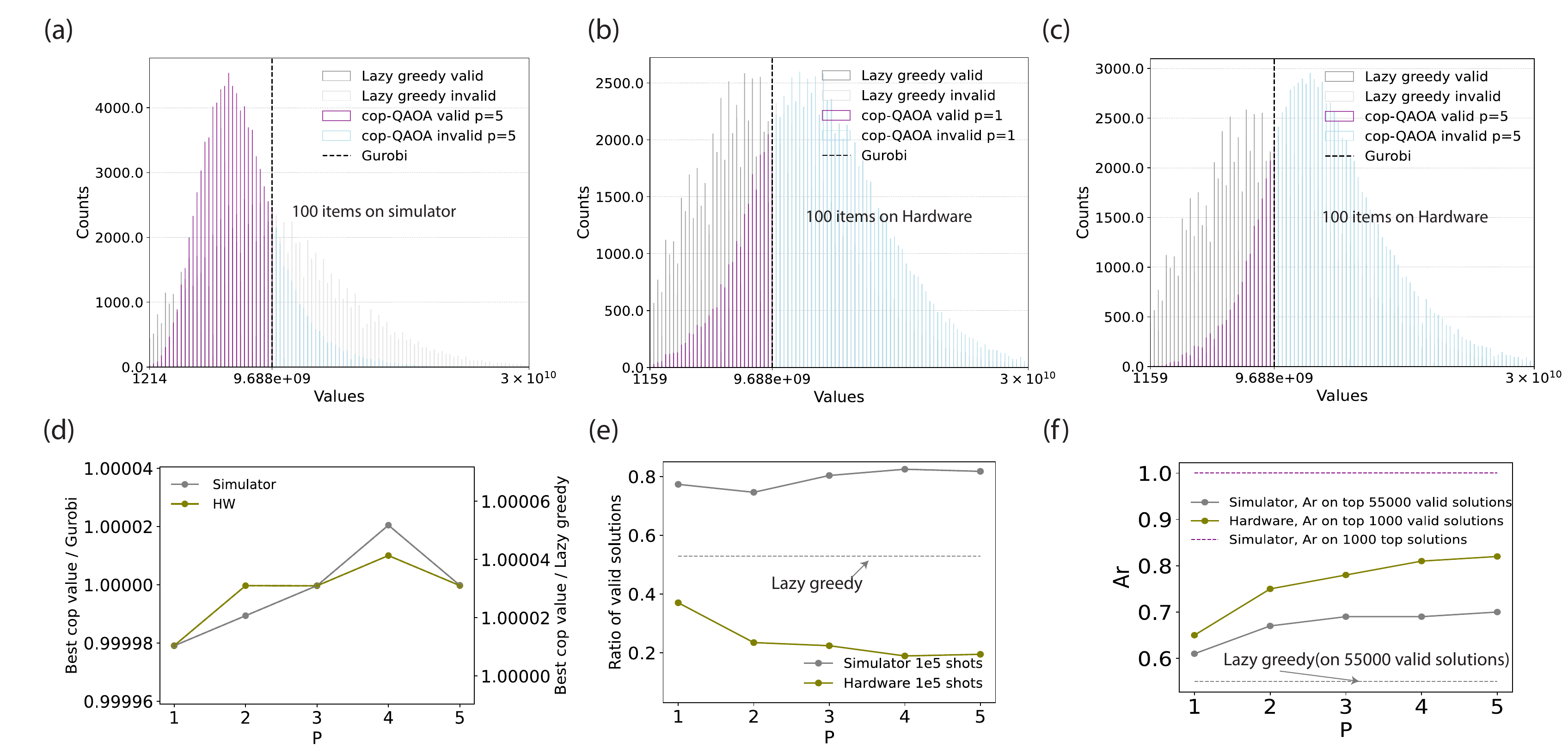}
    \caption{\textbf{Results on 100-item instance}: Cost distribution sampled from lazy greedy  and cop-QAOA (a) simulator, $p=5$; (b) hardware, $p=1$; (c) hardware, $p=5$; (d) the ratio between the best value found by cop-QAOA at each depth $p$ and the solutions obtained by Gurobi and by the lazy greedy; (e) Ratio of valid solutions versus $p$ on simulator and hardware. The horizontal  gray dashed line shows the ratio for lazy greedy; (f) Approximation ratio (Ar) versus $p$ on simulator and hardware. The gray horizontal dashed line shows the approximation ration of lazy greedy. The instance is generated based on  \cite{jooken2022new}. Gurobi reports a solution with an optimality gap of $4 \cdot 10^{-2}$. A random
sampler fails to produce even a single valid solution for
this instance. The number of shots for both the lazy greedy baseline and cop-QAOA is $10^{5}$. The hardware experiments are executed on the \texttt{ibm\_kingston}  quantum processor using the Q-CTRL Performance Management Qiskit function \cite{fire_opal, qiskit-function}. The two-qubit gate depths for $p=1$ through $p=5$ are 5,7,7,5,6. The parameter $k$ for lazy greedy distribution (see Eq. \ref{EQ:proba}) is set to 8.
   }
    \label{fig:fig2}
\end{figure*}
\subsection{Error Mitigation}
The hardware experiments reported in this paper were run on \texttt{ibm\_kingston}, a Heron \texttt{r2} processor with a heavy-hex qubit layout. For these experiments, we made use of the Q-CTRL Performance Management Qiskit function \cite{fire_opal, sachdeva2024quantum, qiskit-function}. Performance management incorporates automated error-suppression techniques, such as layout optimization \cite{hartnett2024learning}, dynamical decoupling for crosstalk and dephasing suppression \cite{ezzell2022dynamical}, and an AI-driven transpilation \cite{fire_opal}. As a point of reference, in Appendix.~\ref{comparison}, we also provide a brief comparison of these results with results obtained on \texttt{ibm\_quebec}, another Heron \texttt{r2} processor, using dynamical decoupling, gate and measurement twirling, without making use of the Performance Management Qiskit function. 
\section{Results}
\begin{figure*}[th]
    \centering
    \includegraphics[width=0.95\linewidth]{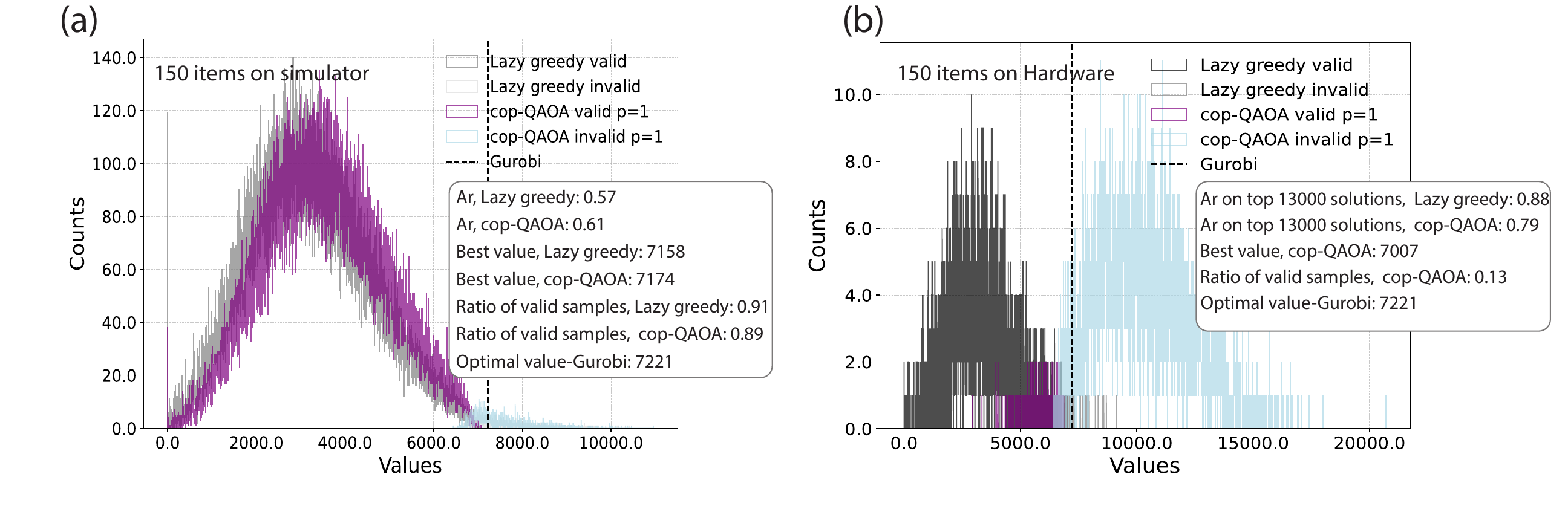}
    \caption{\textbf{Results on 150-item instance}: Cost distribution sampled from lazy greedy and cop-QAOA (a) simulator, $p=1$; (b) hardware, $p=1$. This instance is generated based on Inversely Strongly Correlated Distribution introduced in \cite{pisinger2005hard}. For this instance the solution found by Gurobi is optimal. A random
sampler fails to produce even a single valid solution for
this instance.  The number of shots for both the lazy greedy  baseline and cop-QAOA is $10^{5}$.  The 2-qubit gate depth here is 28. The hardware experiments are executed on the \texttt{ibm\_kingston}  quantum processor using the Q-CTRL Performance Management Qiskit function \cite{fire_opal,qiskit-function}. The parameter $k$ for lazy greedy distribution (see Eq. \ref{EQ:proba}) is set to 10. }
    \label{fig:fig3}
\end{figure*}

In this section, we present the results of our benchmarking experiments. Figures \ref{fig:fig2} and \ref{fig:fig3} show the outcomes for two instances with 100 and 150 items —  both on simulator and IBM hardware — that we selected from the many instances we generated. 

The first instance, with 100 items, is generated based on  the method introduced in \cite{jooken2022new}  and is one of the  cases
where Gurobi struggles to certify optimality. The second
instance, with 150 items, is generated using the Inversely
Strongly Correlated Distribution introduced in \ref{subsec:instance}.  For
this instance, Gurobi successfully finds and verifies the
optimal solution.

We recall that, in cop-QAOA, the initial quantum state is constructed using a smoothed lazy greedy solution as a warm start. Our simulated results reported herein were obtained using the Matrix Product State (MPS) based \texttt{SamplerV2} primitive of \texttt{Qiskit-aer}, (v0.17.1) \cite{qiskit2024}, with no truncation of the bond dimension. In both of these experiments, the number of shots is set to $10^{5}$.

\textbf{100-qubit instance}
Panels (a), (b), and (c) of Fig.~\ref{fig:fig2} show the cost (value) distributions of the solutions found by lazy greedy and cop-QAOA, with $p=5$ on the simulator and $p=1,5$ on hardware, for this instance. In panel (d), the ratios of the best cop-QAOA solution to (i) the solution found by Gurobi and (ii) the best solution found by lazy greedy, are plotted.  Panel (e) shows the ratio of valid solutions found by cop-QAOA, at different values of $p$, in comparison to lazy greedy. While the ratio of valid solutions increases with increasing $p$ on simulator, it decreases with increasing $p$ on hardware. 

Panel (f) shows the approximation ratio versus $p$, both on the simulator and the hardware. Because the fraction of valid samples on hardware decreases with increasing $p$, to ensure a fair comparison, we compute the approximation ratio on hardware for all values of $p$ using only the top 1000  solutions. For the simulator, we report the approximation ratio both for the top 1000 samples for a fair comparison with hardware and for top 55000 valid samples, since lazy greedy finds only 55000 valid solutions. It is evident that on both hardware and simulator the approximation ratio improves with $p$.
Overall, in terms of the best-found  value, cop-QAOA exhibits a slight improvement over solutions obtained by the lazy greedy baseline. This small improvement, which is still remarkable, should be interpreted in light of the strong performance of the warm-start (lazy greedy) solutions, which already achieve values close to the best values observed across all evaluated methods. Consequently, additional gains are necessarily limited. However, when considering the approximation ratio, cop-QAOA clearly outperforms lazy greedy on both the simulator and hardware as there is more room for improvement. We emphasize that a random sampler fails to produce even a single valid solution for this instance.

Note that for this instance, we allowed Gurobi to run for over 70 minutes yet it could not improve upon the solution reported here. In contrast, the total runtime on hardware—including warm-start preparation but excluding classical parameter optimization—remains under 3 minutes for all five QAOA rounds reported. The training time per each run of cop-QAOA is about 2 minutes (with a total of about 40 minutes across the 20 training runs from which we select the best parameters).

\textbf{150-qubit instance}
The 150-qubit instance was generated using the Inversely Strongly Correlated Distribution introduced in  \ref{subsec:instance}. Fig. \ref{fig:fig3} shows the cost distribution of lazy greedy and cop-QAOA with $p=1$ on simulator and  hardware for this instance.  On a noiseless simulator, cop-QAOA exhibits a modest  improvement over lazy greedy, with the approximation ratio increasing from 0.57 to 0.61 and the best value improving from 7158 to 7174. 


On hardware, however, this improvement is not preserved. While cop-QAOA continues to generate valid solutions, the fraction of valid samples drops sharply from 0.89 to 0.13, accompanied by a decrease in  the best observed value (from 7174 to 7007).  We note that for this instance, random sampling fails to produce a single valid solution.

\begin{figure}[h]
    \centering
    \includegraphics[width=0.89\linewidth]{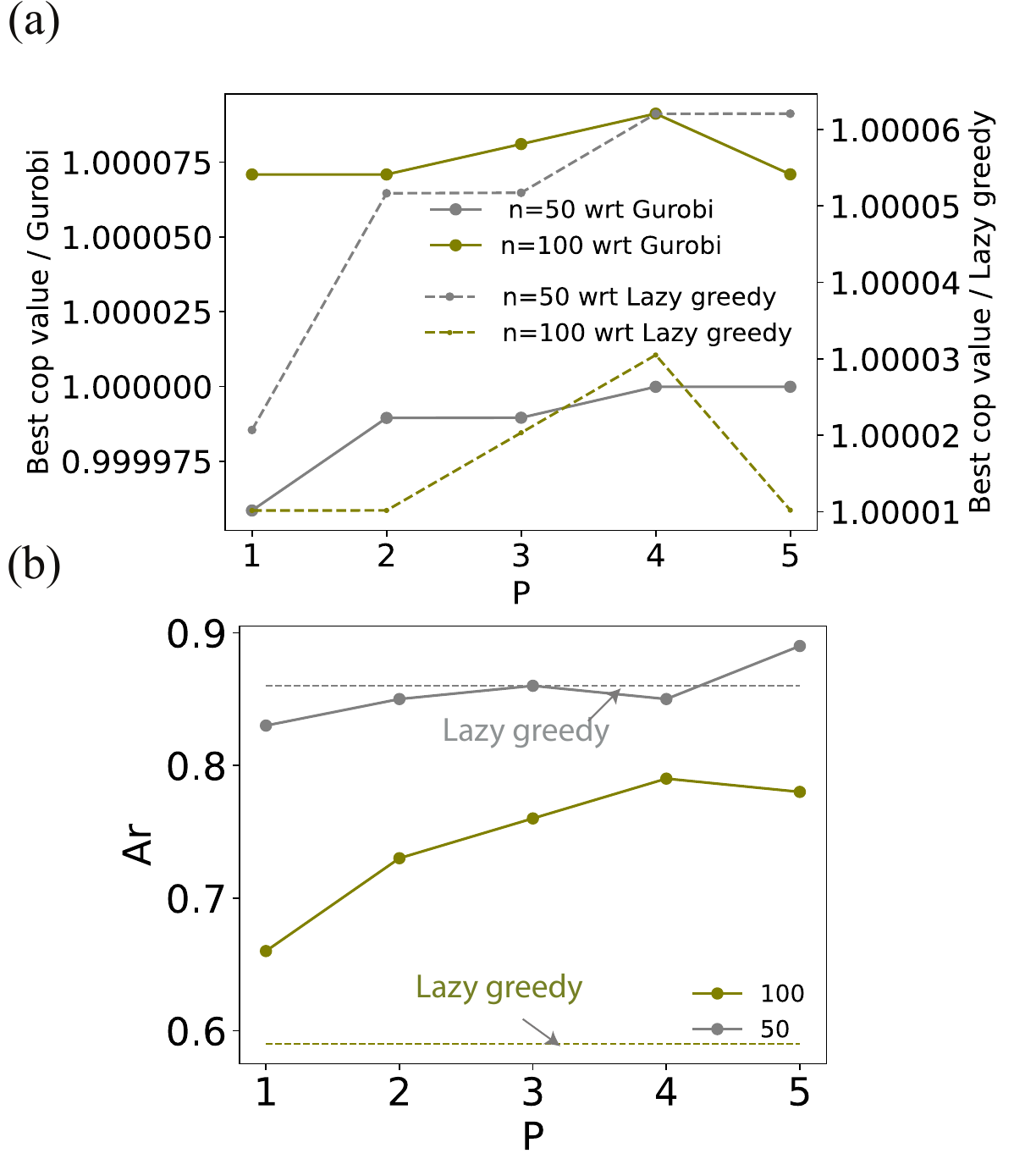}
    \caption{\textbf{Best value and approximation ratio on simulator.} The plot addresses two instances generated based on \cite{jooken2022new}, whose difficulty for Gurobi is discussed in the main text. (a) shows the ratio of the best value found by cop-QAOA to the solutions obtained by Gurobi and by the lazy greedy algorithm, as a function of $p$. (b) shows the approximation ratio as a function of $p$. Horizontal dashed lines show the approximation ratio for lazy greedy.}
    \label{fig:fig5}
\end{figure}

\textbf{Additional instances on simulator} In Fig. \ref{fig:fig5}, we report  results only on simulator for two additional instances with 50 and 100 items generated following method introduced in \cite{jooken2022new}. For each instance, we plot the ratio of the best value obtained by cop-QAOA to those obtained by Gurobi and the lazy greedy algorithm as a function of $p$. It is evident that cop-QAOA can go beyond lazy greedy and even Gurobi. The improvement in terms of best value is modest, which we attribute to the high quality of the warm-start solutions, leaving limited room for further improvement by cop-QAOA.

For the 50-item instance, Gurobi identifies a solution with an optimality gap of $10^{-4}$ within four minutes. While this is very close to optimal, further improvements cannot be achieved within a few minutes longer time limit. For the 100-item instance (shown in green), Gurobi finds a solution within seconds with an optimality gap of $10^{-1}$, but extended a few minutes runtime does not lead to further improvement.   We emphasize that an instance being hard for Gurobi does not necessarily imply that it is universally hard for other solvers.

\subsection{Remarks on the training landscape}
\begin{figure}[h]
    \centering
    \includegraphics[width=0.90\linewidth]{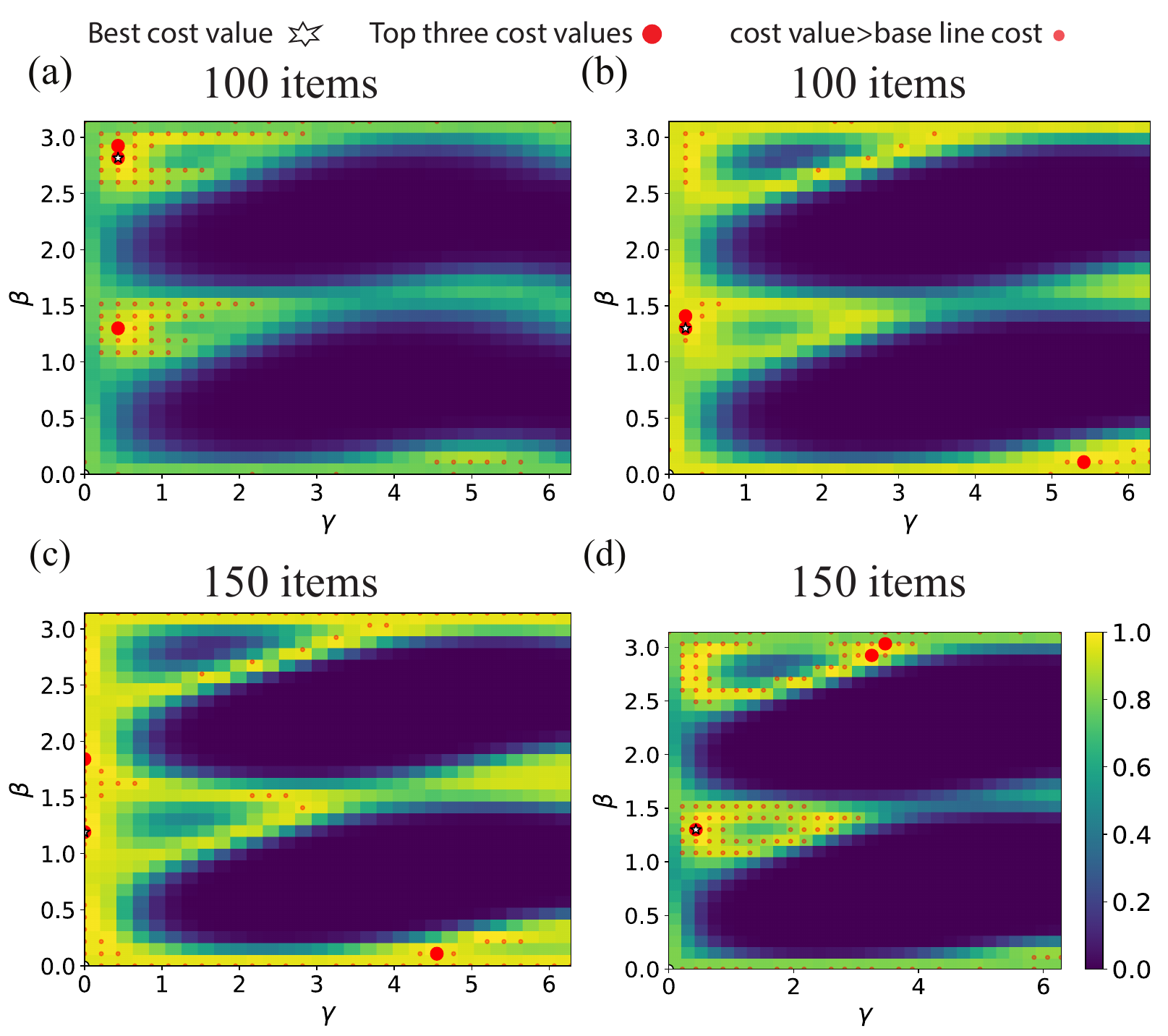}
    \caption{\textbf{Cost value as a function of $\gamma$ and $\beta$ for four problem instances with $P=1$}. Panels (a) and (c) correspond to the same instances shown in Fig. \ref{fig:fig2} and Fig.  \ref{fig:fig3}, respectively. The remaining two instances are generated using the Inversely Strongly Correlated Distribution. Small red dots indicate all $(\gamma, \beta)$ pairs for which the obtained cost exceeds the cost at $(\gamma=0, \beta=0)$. The star marks the best-performing parameter setting, while the large red circles highlight the three best solutions. The similarity of the parameter landscapes—particularly in the regions corresponding to near-optimal values—across different problem instances highlights the potential for parameter transferability and suggests that effective performance may be achieved without instance-specific training. }
    \label{fig:fig4}
\end{figure}

In Fig. \ref{fig:fig4}, we present heat maps of the cost function as a function of $\gamma$ and $\beta$ for several problem instances at qaoa-depth $p=1$. Small red dots indicate all $(\gamma, \beta)$ pairs for which the obtained cost is higher than the cost at $(\gamma=0, \beta=0)$, which corresponds to the lazy greedy  solution. The presence of such points across all instances demonstrates that, by appropriately optimizing the QAOA parameters, one can consistently obtain solutions that improve upon the lazy greedy baseline even at $p=1$. 


Another interesting aspect of this figure is the similarity observed in the parameter landscapes across different problem instances, particularly in the regions where the highest cost values are attained. This structural resemblance suggests that well-performing variational parameters may generalize across instances, supporting the possibility of reusing parameters without instance-specific training and thereby mitigating issues related to barren plateaus.

\section{Conclusion}
In this work, we explore the use of a constrained-optimization variant of the QAOA for the UC problem. We focus on a simplified single-period formulation that preserves the essential combinatorial structure of UC and admits a reduction to the one-dimensional knapsack problem. This setting provides a meaningful and practically motivated benchmark for evaluating near-term quantum optimization methods under realistic constraints.

We applied cop-QAOA, a variant of QAOA that biases the quantum state toward feasible solutions without strictly enforcing feasibility. This approach avoids the circuit depth overhead associated with feasibility-preserving mixers and is therefore well suited to existing quantum hardware. Our numerical and hardware experiments, performed at utility scale with up to 150 qubits, demonstrate that cop-QAOA can consistently improve upon warm-start solutions. Moreover, on selected instances that are challenging for state-of-the-art classical solvers such as Gurobi, it shows the potential to identify slightly better solutions than Gurobi.

Looking forward, this work highlights several promising directions for future research, including extending the approach to multi-period UC with time-coupling constraints, developing improved parameter-transfer and training strategies in particular for constrained optimization, and refining benchmark instance generation to better reflect real-world operational challenges.

\section{Acknowledgements}
We acknowledge use of access provided by Plateforme d' Innovation Numérique et Quantique ($\mathrm{PINQ}^2$) to the \texttt{ibm\_quebec} device, which was used to enable algorithm development and hardware executions in this project. We thank Daniel Egger, Pedro Rivero, and Stefan Woerner for useful discussions on the running of the large scale experiments. We also thank the members of the Quantum Computing for Sustainability Working Group for stimulating discussions.
\bibliography{bibliography}

\begin{thebibliography}{24}%
\makeatletter
\providecommand \@ifxundefined [1]{%
 \@ifx{#1\undefined}
}%
\providecommand \@ifnum [1]{%
 \ifnum #1\expandafter \@firstoftwo
 \else \expandafter \@secondoftwo
 \fi
}%
\providecommand \@ifx [1]{%
 \ifx #1\expandafter \@firstoftwo
 \else \expandafter \@secondoftwo
 \fi
}%
\providecommand \natexlab [1]{#1}%
\providecommand \enquote  [1]{``#1''}%
\providecommand \bibnamefont  [1]{#1}%
\providecommand \bibfnamefont [1]{#1}%
\providecommand \citenamefont [1]{#1}%
\providecommand \href@noop [0]{\@secondoftwo}%
\providecommand \href [0]{\begingroup \@sanitize@url \@href}%
\providecommand \@href[1]{\@@startlink{#1}\@@href}%
\providecommand \@@href[1]{\endgroup#1\@@endlink}%
\providecommand \@sanitize@url [0]{\catcode `\\12\catcode `\$12\catcode
  `\&12\catcode `\#12\catcode `\^12\catcode `\_12\catcode `\%12\relax}%
\providecommand \@@startlink[1]{}%
\providecommand \@@endlink[0]{}%
\providecommand \url  [0]{\begingroup\@sanitize@url \@url }%
\providecommand \@url [1]{\endgroup\@href {#1}{\urlprefix }}%
\providecommand \urlprefix  [0]{URL }%
\providecommand \Eprint [0]{\href }%
\providecommand \doibase [0]{http://dx.doi.org/}%
\providecommand \selectlanguage [0]{\@gobble}%
\providecommand \bibinfo  [0]{\@secondoftwo}%
\providecommand \bibfield  [0]{\@secondoftwo}%
\providecommand \translation [1]{[#1]}%
\providecommand \BibitemOpen [0]{}%
\providecommand \bibitemStop [0]{}%
\providecommand \bibitemNoStop [0]{.\EOS\space}%
\providecommand \EOS [0]{\spacefactor3000\relax}%
\providecommand \BibitemShut  [1]{\csname bibitem#1\endcsname}%
\let\auto@bib@innerbib\@empty
\bibitem [{\citenamefont {Agliardi}\ \emph {et~al.}(2024)\citenamefont
  {Agliardi}, \citenamefont {Cortiana}, \citenamefont {Dekusar}, \citenamefont
  {Ghosh}, \citenamefont {Mohseni}, \citenamefont {O'Meara}, \citenamefont
  {Valls}, \citenamefont {Yogaraj},\ and\ \citenamefont
  {Zhuk}}]{agliardi2024machine}%
  \BibitemOpen
  \bibfield  {author} {\bibinfo {author} {\bibfnamefont {Gabriele}\
  \bibnamefont {Agliardi}}, \bibinfo {author} {\bibfnamefont {Giorgio}\
  \bibnamefont {Cortiana}}, \bibinfo {author} {\bibfnamefont {Anton}\
  \bibnamefont {Dekusar}}, \bibinfo {author} {\bibfnamefont {Kumar}\
  \bibnamefont {Ghosh}}, \bibinfo {author} {\bibfnamefont {Naeimeh}\
  \bibnamefont {Mohseni}}, \bibinfo {author} {\bibfnamefont {Corey}\
  \bibnamefont {O'Meara}}, \bibinfo {author} {\bibfnamefont {V{\'\i}ctor}\
  \bibnamefont {Valls}}, \bibinfo {author} {\bibfnamefont {Kavitha}\
  \bibnamefont {Yogaraj}}, \ and\ \bibinfo {author} {\bibfnamefont {Sergiy}\
  \bibnamefont {Zhuk}},\ }\bibfield  {title} {\enquote {\bibinfo {title} {A
  machine learning approach to boost the vehicle-2-grid scheduling},}\ }in\
  \href@noop {} {\emph {\bibinfo {booktitle} {2024 IEEE Sustainable Power and
  Energy Conference (iSPEC)}}}\ (\bibinfo {organization} {IEEE},\ \bibinfo
  {year} {2024})\ pp.\ \bibinfo {pages} {670--675}\BibitemShut {NoStop}%
\bibitem [{\citenamefont {Mohseni}\ \emph {et~al.}(2025)\citenamefont
  {Mohseni}, \citenamefont {Morstyn}, \citenamefont {O’Meara}, \citenamefont
  {Bucher}, \citenamefont {N{\"u}{\ss}lein},\ and\ \citenamefont
  {Cortiana}}]{mohseni2025evidence}%
  \BibitemOpen
  \bibfield  {author} {\bibinfo {author} {\bibfnamefont {Naeimeh}\ \bibnamefont
  {Mohseni}}, \bibinfo {author} {\bibfnamefont {Thomas}\ \bibnamefont
  {Morstyn}}, \bibinfo {author} {\bibfnamefont {Corey}\ \bibnamefont
  {O’Meara}}, \bibinfo {author} {\bibfnamefont {David}\ \bibnamefont
  {Bucher}}, \bibinfo {author} {\bibfnamefont {Jonas}\ \bibnamefont
  {N{\"u}{\ss}lein}}, \ and\ \bibinfo {author} {\bibfnamefont {Giorgio}\
  \bibnamefont {Cortiana}},\ }\bibfield  {title} {\enquote {\bibinfo {title}
  {Evidence of quantum scaling advantage in approximate optimization for energy
  coalition formation with 100+ agents},}\ }\href@noop {} {\bibfield  {journal}
  {\bibinfo  {journal} {Quantum Science and Technology}\ }\textbf {\bibinfo
  {volume} {11}},\ \bibinfo {pages} {015009} (\bibinfo {year}
  {2025})}\BibitemShut {NoStop}%
\bibitem [{\citenamefont {Koretsky}\ \emph {et~al.}(2021)\citenamefont
  {Koretsky}, \citenamefont {Gokhale}, \citenamefont {Baker}, \citenamefont
  {Viszlai}, \citenamefont {Zheng}, \citenamefont {Gurung}, \citenamefont
  {Burg}, \citenamefont {Paaso}, \citenamefont {Khodaei}, \citenamefont
  {Eskandarpour} \emph {et~al.}}]{koretsky2021adapting}%
  \BibitemOpen
  \bibfield  {author} {\bibinfo {author} {\bibfnamefont {Samantha}\
  \bibnamefont {Koretsky}}, \bibinfo {author} {\bibfnamefont {Pranav}\
  \bibnamefont {Gokhale}}, \bibinfo {author} {\bibfnamefont {Jonathan~M}\
  \bibnamefont {Baker}}, \bibinfo {author} {\bibfnamefont {Joshua}\
  \bibnamefont {Viszlai}}, \bibinfo {author} {\bibfnamefont {Honghao}\
  \bibnamefont {Zheng}}, \bibinfo {author} {\bibfnamefont {Niroj}\ \bibnamefont
  {Gurung}}, \bibinfo {author} {\bibfnamefont {Ryan}\ \bibnamefont {Burg}},
  \bibinfo {author} {\bibfnamefont {Esa~Aleksi}\ \bibnamefont {Paaso}},
  \bibinfo {author} {\bibfnamefont {Amin}\ \bibnamefont {Khodaei}}, \bibinfo
  {author} {\bibfnamefont {Rozhin}\ \bibnamefont {Eskandarpour}},  \emph
  {et~al.},\ }\bibfield  {title} {\enquote {\bibinfo {title} {Adapting quantum
  approximation optimization algorithm (qaoa) for unit commitment},}\ }in\
  \href@noop {} {\emph {\bibinfo {booktitle} {2021 IEEE International
  Conference on Quantum Computing and Engineering (QCE)}}}\ (\bibinfo
  {organization} {IEEE},\ \bibinfo {year} {2021})\ pp.\ \bibinfo {pages}
  {181--187}\BibitemShut {NoStop}%
\bibitem [{\citenamefont {Agliardi}\ \emph {et~al.}(2025)\citenamefont
  {Agliardi}, \citenamefont {Cortiana}, \citenamefont {Dekusar}, \citenamefont
  {Ghosh}, \citenamefont {Mohseni}, \citenamefont {O’Meara}, \citenamefont
  {Valls}, \citenamefont {Yogaraj},\ and\ \citenamefont
  {Zhuk}}]{agliardi2025mitigating}%
  \BibitemOpen
  \bibfield  {author} {\bibinfo {author} {\bibfnamefont {Gabriele}\
  \bibnamefont {Agliardi}}, \bibinfo {author} {\bibfnamefont {Giorgio}\
  \bibnamefont {Cortiana}}, \bibinfo {author} {\bibfnamefont {Anton}\
  \bibnamefont {Dekusar}}, \bibinfo {author} {\bibfnamefont {Kumar}\
  \bibnamefont {Ghosh}}, \bibinfo {author} {\bibfnamefont {Naeimeh}\
  \bibnamefont {Mohseni}}, \bibinfo {author} {\bibfnamefont {Corey}\
  \bibnamefont {O’Meara}}, \bibinfo {author} {\bibfnamefont {V{\'\i}ctor}\
  \bibnamefont {Valls}}, \bibinfo {author} {\bibfnamefont {Kavitha}\
  \bibnamefont {Yogaraj}}, \ and\ \bibinfo {author} {\bibfnamefont {Sergiy}\
  \bibnamefont {Zhuk}},\ }\bibfield  {title} {\enquote {\bibinfo {title}
  {Mitigating exponential concentration in covariant quantum kernels for
  subspace and real-world data},}\ }\href@noop {} {\bibfield  {journal}
  {\bibinfo  {journal} {npj Quantum Information}\ } (\bibinfo {year}
  {2025})}\BibitemShut {NoStop}%
\bibitem [{\citenamefont {Hadfield}\ \emph {et~al.}(2019)\citenamefont
  {Hadfield}, \citenamefont {Wang}, \citenamefont {O’gorman}, \citenamefont
  {Rieffel}, \citenamefont {Venturelli},\ and\ \citenamefont
  {Biswas}}]{hadfield2019quantum}%
  \BibitemOpen
  \bibfield  {author} {\bibinfo {author} {\bibfnamefont {Stuart}\ \bibnamefont
  {Hadfield}}, \bibinfo {author} {\bibfnamefont {Zhihui}\ \bibnamefont {Wang}},
  \bibinfo {author} {\bibfnamefont {Bryan}\ \bibnamefont {O’gorman}},
  \bibinfo {author} {\bibfnamefont {Eleanor~G}\ \bibnamefont {Rieffel}},
  \bibinfo {author} {\bibfnamefont {Davide}\ \bibnamefont {Venturelli}}, \ and\
  \bibinfo {author} {\bibfnamefont {Rupak}\ \bibnamefont {Biswas}},\ }\bibfield
   {title} {\enquote {\bibinfo {title} {From the quantum approximate
  optimization algorithm to a quantum alternating operator ansatz},}\
  }\href@noop {} {\bibfield  {journal} {\bibinfo  {journal} {Algorithms}\
  }\textbf {\bibinfo {volume} {12}},\ \bibinfo {pages} {34} (\bibinfo {year}
  {2019})}\BibitemShut {NoStop}%
\bibitem [{\citenamefont {Wang}\ \emph {et~al.}(2020)\citenamefont {Wang},
  \citenamefont {Rubin}, \citenamefont {Dominy},\ and\ \citenamefont
  {Rieffel}}]{wang2020xy}%
  \BibitemOpen
  \bibfield  {author} {\bibinfo {author} {\bibfnamefont {Zhihui}\ \bibnamefont
  {Wang}}, \bibinfo {author} {\bibfnamefont {Nicholas~C}\ \bibnamefont
  {Rubin}}, \bibinfo {author} {\bibfnamefont {Jason~M}\ \bibnamefont {Dominy}},
  \ and\ \bibinfo {author} {\bibfnamefont {Eleanor~G}\ \bibnamefont
  {Rieffel}},\ }\bibfield  {title} {\enquote {\bibinfo {title} {Xy mixers:
  Analytical and numerical results for the quantum alternating operator
  ansatz},}\ }\href@noop {} {\bibfield  {journal} {\bibinfo  {journal}
  {Physical Review A}\ }\textbf {\bibinfo {volume} {101}},\ \bibinfo {pages}
  {012320} (\bibinfo {year} {2020})}\BibitemShut {NoStop}%
\bibitem [{\citenamefont {B{\"a}rtschi}\ and\ \citenamefont
  {Eidenbenz}(2020)}]{bartschi2020grover}%
  \BibitemOpen
  \bibfield  {author} {\bibinfo {author} {\bibfnamefont {Andreas}\ \bibnamefont
  {B{\"a}rtschi}}\ and\ \bibinfo {author} {\bibfnamefont {Stephan}\
  \bibnamefont {Eidenbenz}},\ }\bibfield  {title} {\enquote {\bibinfo {title}
  {Grover mixers for qaoa: Shifting complexity from mixer design to state
  preparation},}\ }in\ \href@noop {} {\emph {\bibinfo {booktitle} {2020 IEEE
  International Conference on Quantum Computing and Engineering (QCE)}}}\
  (\bibinfo {organization} {IEEE},\ \bibinfo {year} {2020})\ pp.\ \bibinfo
  {pages} {72--82}\BibitemShut {NoStop}%
\bibitem [{\citenamefont {Van~Dam}\ \emph {et~al.}(2021)\citenamefont
  {Van~Dam}, \citenamefont {Eldefrawy}, \citenamefont {Genise},\ and\
  \citenamefont {Parham}}]{van2021quantum}%
  \BibitemOpen
  \bibfield  {author} {\bibinfo {author} {\bibfnamefont {Wim}\ \bibnamefont
  {Van~Dam}}, \bibinfo {author} {\bibfnamefont {Karim}\ \bibnamefont
  {Eldefrawy}}, \bibinfo {author} {\bibfnamefont {Nicholas}\ \bibnamefont
  {Genise}}, \ and\ \bibinfo {author} {\bibfnamefont {Natalie}\ \bibnamefont
  {Parham}},\ }\bibfield  {title} {\enquote {\bibinfo {title} {Quantum
  optimization heuristics with an application to knapsack problems},}\ }in\
  \href@noop {} {\emph {\bibinfo {booktitle} {2021 IEEE International
  Conference on Quantum Computing and Engineering (QCE)}}}\ (\bibinfo
  {organization} {IEEE},\ \bibinfo {year} {2021})\ pp.\ \bibinfo {pages}
  {160--170}\BibitemShut {NoStop}%
\bibitem [{\citenamefont {Barkoutsos}\ \emph {et~al.}(2020)\citenamefont
  {Barkoutsos}, \citenamefont {Nannicini}, \citenamefont {Robert},
  \citenamefont {Tavernelli},\ and\ \citenamefont
  {Woerner}}]{Barkoutsos2020improving}%
  \BibitemOpen
  \bibfield  {author} {\bibinfo {author} {\bibfnamefont {Panagiotis~Kl.}\
  \bibnamefont {Barkoutsos}}, \bibinfo {author} {\bibfnamefont {Giacomo}\
  \bibnamefont {Nannicini}}, \bibinfo {author} {\bibfnamefont {Anton}\
  \bibnamefont {Robert}}, \bibinfo {author} {\bibfnamefont {Ivano}\
  \bibnamefont {Tavernelli}}, \ and\ \bibinfo {author} {\bibfnamefont {Stefan}\
  \bibnamefont {Woerner}},\ }\bibfield  {title} {\enquote {\bibinfo {title}
  {Improving {V}ariational {Q}uantum {O}ptimization using {CV}a{R}},}\ }\href
  {\doibase 10.22331/q-2020-04-20-256} {\bibfield  {journal} {\bibinfo
  {journal} {{Quantum}}\ }\textbf {\bibinfo {volume} {4}},\ \bibinfo {pages}
  {256} (\bibinfo {year} {2020})}\BibitemShut {NoStop}%
\bibitem [{\citenamefont {Bendotti}\ \emph {et~al.}(2019)\citenamefont
  {Bendotti}, \citenamefont {Fouilhoux},\ and\ \citenamefont
  {Rottner}}]{bendotti2019complexity}%
  \BibitemOpen
  \bibfield  {author} {\bibinfo {author} {\bibfnamefont {Pascale}\ \bibnamefont
  {Bendotti}}, \bibinfo {author} {\bibfnamefont {Pierre}\ \bibnamefont
  {Fouilhoux}}, \ and\ \bibinfo {author} {\bibfnamefont {C{\'e}cile}\
  \bibnamefont {Rottner}},\ }\bibfield  {title} {\enquote {\bibinfo {title} {On
  the complexity of the unit commitment problem},}\ }\href@noop {} {\bibfield
  {journal} {\bibinfo  {journal} {Annals of Operations Research}\ }\textbf
  {\bibinfo {volume} {274}},\ \bibinfo {pages} {119--130} (\bibinfo {year}
  {2019})}\BibitemShut {NoStop}%
\bibitem [{\citenamefont {{Gurobi Optimization, LLC}}(2024)}]{gurobi}%
  \BibitemOpen
  \bibfield  {author} {\bibinfo {author} {\bibnamefont {{Gurobi Optimization,
  LLC}}},\ }\href {https://www.gurobi.com} {\enquote {\bibinfo {title} {{Gurobi
  Optimizer Reference Manual}},}\ } (\bibinfo {year} {2024})\BibitemShut
  {NoStop}%
\bibitem [{\citenamefont {Farhi}\ \emph {et~al.}(2014)\citenamefont {Farhi},
  \citenamefont {Goldstone},\ and\ \citenamefont {Gutmann}}]{farhi2014qaoa}%
  \BibitemOpen
  \bibfield  {author} {\bibinfo {author} {\bibfnamefont {Edward}\ \bibnamefont
  {Farhi}}, \bibinfo {author} {\bibfnamefont {Jeffrey}\ \bibnamefont
  {Goldstone}}, \ and\ \bibinfo {author} {\bibfnamefont {Sam}\ \bibnamefont
  {Gutmann}},\ }\bibfield  {title} {\enquote {\bibinfo {title} {A quantum
  approximate optimization algorithm},}\ }\href
  {https://arxiv.org/abs/1411.4028} {\bibfield  {journal} {\bibinfo  {journal}
  {arXiv preprint arXiv:1411.4028}\ } (\bibinfo {year} {2014})}\BibitemShut
  {NoStop}%
\bibitem [{\citenamefont {Christiansen}\ \emph {et~al.}(2024)\citenamefont
  {Christiansen}, \citenamefont {Binkowski}, \citenamefont {Ramacciotti},\ and\
  \citenamefont {Wilkening}}]{christiansen2024quantum}%
  \BibitemOpen
  \bibfield  {author} {\bibinfo {author} {\bibfnamefont {Paul}\ \bibnamefont
  {Christiansen}}, \bibinfo {author} {\bibfnamefont {Lennart}\ \bibnamefont
  {Binkowski}}, \bibinfo {author} {\bibfnamefont {Debora}\ \bibnamefont
  {Ramacciotti}}, \ and\ \bibinfo {author} {\bibfnamefont {S{\"o}ren}\
  \bibnamefont {Wilkening}},\ }\bibfield  {title} {\enquote {\bibinfo {title}
  {Quantum tree generator improves qaoa state-of-the-art for the knapsack
  problem},}\ }\href@noop {} {\bibfield  {journal} {\bibinfo  {journal} {arXiv
  preprint arXiv:2411.00518}\ } (\bibinfo {year} {2024})}\BibitemShut {NoStop}%
\bibitem [{\citenamefont {Pisinger}(2005)}]{pisinger2005hard}%
  \BibitemOpen
  \bibfield  {author} {\bibinfo {author} {\bibfnamefont {David}\ \bibnamefont
  {Pisinger}},\ }\bibfield  {title} {\enquote {\bibinfo {title} {Where are the
  hard knapsack problems?}}\ }\href@noop {} {\bibfield  {journal} {\bibinfo
  {journal} {Computers \& Operations Research}\ }\textbf {\bibinfo {volume}
  {32}},\ \bibinfo {pages} {2271--2284} (\bibinfo {year} {2005})}\BibitemShut
  {NoStop}%
\bibitem [{\citenamefont {Jooken}\ \emph {et~al.}(2022)\citenamefont {Jooken},
  \citenamefont {Leyman},\ and\ \citenamefont
  {De~Causmaecker}}]{jooken2022new}%
  \BibitemOpen
  \bibfield  {author} {\bibinfo {author} {\bibfnamefont {Jorik}\ \bibnamefont
  {Jooken}}, \bibinfo {author} {\bibfnamefont {Pieter}\ \bibnamefont {Leyman}},
  \ and\ \bibinfo {author} {\bibfnamefont {Patrick}\ \bibnamefont
  {De~Causmaecker}},\ }\bibfield  {title} {\enquote {\bibinfo {title} {A new
  class of hard problem instances for the 0--1 knapsack problem},}\ }\href@noop
  {} {\bibfield  {journal} {\bibinfo  {journal} {European Journal of
  Operational Research}\ }\textbf {\bibinfo {volume} {301}},\ \bibinfo {pages}
  {841--854} (\bibinfo {year} {2022})}\BibitemShut {NoStop}%
\bibitem [{\citenamefont {Barron}\ \emph {et~al.}(2024)\citenamefont {Barron},
  \citenamefont {Egger}, \citenamefont {Pelofske}, \citenamefont
  {B{\"a}rtschi}, \citenamefont {Eidenbenz}, \citenamefont {Lehmkuehler},\ and\
  \citenamefont {Woerner}}]{barron2024provable}%
  \BibitemOpen
  \bibfield  {author} {\bibinfo {author} {\bibfnamefont {Samantha~V}\
  \bibnamefont {Barron}}, \bibinfo {author} {\bibfnamefont {Daniel~J}\
  \bibnamefont {Egger}}, \bibinfo {author} {\bibfnamefont {Elijah}\
  \bibnamefont {Pelofske}}, \bibinfo {author} {\bibfnamefont {Andreas}\
  \bibnamefont {B{\"a}rtschi}}, \bibinfo {author} {\bibfnamefont {Stephan}\
  \bibnamefont {Eidenbenz}}, \bibinfo {author} {\bibfnamefont {Matthis}\
  \bibnamefont {Lehmkuehler}}, \ and\ \bibinfo {author} {\bibfnamefont
  {Stefan}\ \bibnamefont {Woerner}},\ }\bibfield  {title} {\enquote {\bibinfo
  {title} {Provable bounds for noise-free expectation values computed from
  noisy samples},}\ }\href@noop {} {\bibfield  {journal} {\bibinfo  {journal}
  {Nature Computational Science}\ }\textbf {\bibinfo {volume} {4}},\ \bibinfo
  {pages} {865--875} (\bibinfo {year} {2024})}\BibitemShut {NoStop}%
\bibitem [{\citenamefont {Q-CTRL}(2025)}]{fire_opal}%
  \BibitemOpen
  \bibfield  {author} {\bibinfo {author} {\bibnamefont {Q-CTRL}},\ }\href@noop
  {} {\enquote {\bibinfo {title} {Fire {O}pal},}\ }\bibinfo {howpublished}
  {https://q-ctrl.com/fire-opal} (\bibinfo {year} {2025}),\ \bibinfo {note}
  {[Online]}\BibitemShut {NoStop}%
\bibitem [{qis(2025)}]{qiskit-function}%
  \BibitemOpen
  \href@noop {} {\enquote {\bibinfo {title} {{Performance Management: A Qiskit
  Function by Q-CTRL Fire Opal}},}\ }\bibinfo {howpublished}
  {\url{https://quantum.cloud.ibm.com/docs/en/guides/q-ctrl-performance-management}}
  (\bibinfo {year} {2025}),\ \bibinfo {note} {accessed: December 01,
  2025}\BibitemShut {NoStop}%
\bibitem [{\citenamefont {Sachdeva}\ \emph {et~al.}(2024)\citenamefont
  {Sachdeva}, \citenamefont {Hartnett}, \citenamefont {Maity}, \citenamefont
  {Marsh}, \citenamefont {Wang}, \citenamefont {Winick}, \citenamefont
  {Dougherty}, \citenamefont {Canuto}, \citenamefont {Chong}, \citenamefont
  {Hush} \emph {et~al.}}]{sachdeva2024quantum}%
  \BibitemOpen
  \bibfield  {author} {\bibinfo {author} {\bibfnamefont {Natasha}\ \bibnamefont
  {Sachdeva}}, \bibinfo {author} {\bibfnamefont {Gavin~S}\ \bibnamefont
  {Hartnett}}, \bibinfo {author} {\bibfnamefont {Smarak}\ \bibnamefont
  {Maity}}, \bibinfo {author} {\bibfnamefont {Samuel}\ \bibnamefont {Marsh}},
  \bibinfo {author} {\bibfnamefont {Yulun}\ \bibnamefont {Wang}}, \bibinfo
  {author} {\bibfnamefont {Adam}\ \bibnamefont {Winick}}, \bibinfo {author}
  {\bibfnamefont {Ryan}\ \bibnamefont {Dougherty}}, \bibinfo {author}
  {\bibfnamefont {Daniel}\ \bibnamefont {Canuto}}, \bibinfo {author}
  {\bibfnamefont {You~Quan}\ \bibnamefont {Chong}}, \bibinfo {author}
  {\bibfnamefont {Michael}\ \bibnamefont {Hush}},  \emph {et~al.},\ }\bibfield
  {title} {\enquote {\bibinfo {title} {Quantum optimization using a 127-qubit
  gate-model ibm quantum computer can outperform quantum annealers for
  nontrivial binary optimization problems},}\ }\href@noop {} {\bibfield
  {journal} {\bibinfo  {journal} {arXiv preprint arXiv:2406.01743}\ } (\bibinfo
  {year} {2024})}\BibitemShut {NoStop}%
\bibitem [{\citenamefont {Hartnett}\ \emph {et~al.}(2024)\citenamefont
  {Hartnett}, \citenamefont {Barbosa}, \citenamefont {Mundada}, \citenamefont
  {Hush}, \citenamefont {Biercuk},\ and\ \citenamefont
  {Baum}}]{hartnett2024learning}%
  \BibitemOpen
  \bibfield  {author} {\bibinfo {author} {\bibfnamefont {Gavin~S}\ \bibnamefont
  {Hartnett}}, \bibinfo {author} {\bibfnamefont {Aaron}\ \bibnamefont
  {Barbosa}}, \bibinfo {author} {\bibfnamefont {Pranav~S}\ \bibnamefont
  {Mundada}}, \bibinfo {author} {\bibfnamefont {Michael}\ \bibnamefont {Hush}},
  \bibinfo {author} {\bibfnamefont {Michael~J}\ \bibnamefont {Biercuk}}, \ and\
  \bibinfo {author} {\bibfnamefont {Yuval}\ \bibnamefont {Baum}},\ }\bibfield
  {title} {\enquote {\bibinfo {title} {Learning to rank quantum circuits for
  hardware-optimized performance enhancement},}\ }\href@noop {} {\bibfield
  {journal} {\bibinfo  {journal} {Quantum}\ }\textbf {\bibinfo {volume} {8}},\
  \bibinfo {pages} {1542} (\bibinfo {year} {2024})}\BibitemShut {NoStop}%
\bibitem [{\citenamefont {Ezzell}\ \emph {et~al.}(2022)\citenamefont {Ezzell},
  \citenamefont {Pokharel}, \citenamefont {Tewala}, \citenamefont {Quiroz},\
  and\ \citenamefont {Lidar}}]{ezzell2022dynamical}%
  \BibitemOpen
  \bibfield  {author} {\bibinfo {author} {\bibfnamefont {Nic}\ \bibnamefont
  {Ezzell}}, \bibinfo {author} {\bibfnamefont {Bibek}\ \bibnamefont
  {Pokharel}}, \bibinfo {author} {\bibfnamefont {Lina}\ \bibnamefont {Tewala}},
  \bibinfo {author} {\bibfnamefont {Gregory}\ \bibnamefont {Quiroz}}, \ and\
  \bibinfo {author} {\bibfnamefont {Daniel~A}\ \bibnamefont {Lidar}},\
  }\bibfield  {title} {\enquote {\bibinfo {title} {Dynamical decoupling for
  superconducting qubits: A performance survey},}\ }\href@noop {} {\bibfield
  {journal} {\bibinfo  {journal} {arXiv preprint arXiv:2207.03670}\ } (\bibinfo
  {year} {2022})}\BibitemShut {NoStop}%
\bibitem [{\citenamefont {Javadi-Abhari}\ \emph {et~al.}(2024)\citenamefont
  {Javadi-Abhari}, \citenamefont {Treinish}, \citenamefont {Krsulich},
  \citenamefont {Wood}, \citenamefont {Lishman}, \citenamefont {Gacon},
  \citenamefont {Martiel}, \citenamefont {Nation}, \citenamefont {Bishop},
  \citenamefont {Cross}, \citenamefont {Johnson},\ and\ \citenamefont
  {Gambetta}}]{qiskit2024}%
  \BibitemOpen
  \bibfield  {author} {\bibinfo {author} {\bibfnamefont {Ali}\ \bibnamefont
  {Javadi-Abhari}}, \bibinfo {author} {\bibfnamefont {Matthew}\ \bibnamefont
  {Treinish}}, \bibinfo {author} {\bibfnamefont {Kevin}\ \bibnamefont
  {Krsulich}}, \bibinfo {author} {\bibfnamefont {Christopher~J.}\ \bibnamefont
  {Wood}}, \bibinfo {author} {\bibfnamefont {Jake}\ \bibnamefont {Lishman}},
  \bibinfo {author} {\bibfnamefont {Julien}\ \bibnamefont {Gacon}}, \bibinfo
  {author} {\bibfnamefont {Simon}\ \bibnamefont {Martiel}}, \bibinfo {author}
  {\bibfnamefont {Paul~D.}\ \bibnamefont {Nation}}, \bibinfo {author}
  {\bibfnamefont {Lev~S.}\ \bibnamefont {Bishop}}, \bibinfo {author}
  {\bibfnamefont {Andrew~W.}\ \bibnamefont {Cross}}, \bibinfo {author}
  {\bibfnamefont {Blake~R.}\ \bibnamefont {Johnson}}, \ and\ \bibinfo {author}
  {\bibfnamefont {Jay~M.}\ \bibnamefont {Gambetta}},\ }\href {\doibase
  10.48550/arXiv.2405.08810} {\enquote {\bibinfo {title} {Quantum computing
  with {Q}iskit},}\ } (\bibinfo {year} {2024}),\ \Eprint
  {http://arxiv.org/abs/2405.08810} {arXiv:2405.08810 [quant-ph]} \BibitemShut
  {NoStop}%
\bibitem [{emd(2025)}]{emdocs}%
  \BibitemOpen
  \href@noop {} {\enquote {\bibinfo {title} {{Error mitigation and suppression
  techniques}},}\ }\bibinfo {howpublished}
  {\url{https://quantum.cloud.ibm.com/docs/en/guides/error-mitigation-and-suppression-techniques}}
  (\bibinfo {year} {2025}),\ \bibinfo {note} {accessed: January 22,
  2026}\BibitemShut {NoStop}%
\bibitem [{fr(2025)}]{fr}%
  \BibitemOpen
  \href@noop {} {\enquote {\bibinfo {title} {{Fractional Gates}},}\ }\bibinfo
  {howpublished}
  {\url{https://quantum.cloud.ibm.com/docs/en/guides/fractional-gates}}
  (\bibinfo {year} {2025}),\ \bibinfo {note} {accessed: January 24,
  2026}\BibitemShut {NoStop}%
\end{thebibliography}%
\appendix
\section{Derivative Trick} \label{app:derivative}

Recall that our goal is to minimize the function

\begin{align}
    C(\vec{y},\vec{p}) := \sum_{i=1}^{n} \left( A_i \cdot y_i + B_i \cdot p_i + C_i \cdot p_i^2 \right),
    \label{eq:UC-objective-app}
\end{align}

subject to the constraints 
\begin{align}
    \sum_{i=1}^{n} p_i &= L, \quad \forall t \quad \text{(demand constraint)}, \label{eq:demand_constraint} \\
    p_{i,\text{min}} \cdot y_i &\leq p_i \leq p_{i,\text{max}} \cdot y_i, \quad \forall i \quad \text{(gen. limits)}, \label{eq:gen_limits} 
\end{align}
with the $y_i$ binary variables and the $p_i$ real valued variables. For any fixed value of the $y_i$ variables, this constrained optimization problem takes its minimum value when the $p_i$ variables satisfy the Karush-Kuhn-Tucker (KKT) criterion:

\begin{align}
\partial \bigg( C(\vec{y}, \vec{p}) &+ \lambda \left(L - \sum_i  p_i \right) + \sum_i \xi_i (y_i p_{i,\min}  - p_i) \nonumber \\
&+ \sum_i \eta_i(p_i - y_i p_{i,\max}) \bigg) \ni 0 \label{eq:opt_1}
\end{align}
such that
\begin{align}
L - \sum_i p_i &\leq 0 \label{eq:opt_2}\\
\lambda &\geq 0 \label{eq:opt_3}\\
\lambda \left(\sum_i p_i - L \right) &= 0 \label{eq:opt_4}
\end{align}
and $\forall i$:
\begin{align}
y_i p_{i,\min} - p_i &\leq 0 \label{eq:opt_5}\\
\xi_i &\geq 0 \label{eq:opt_6}\\
\xi_i (p_i - y_i p_{i,\min}) &= 0 \label{eq:opt_7}\\
p_i - y_i p_{i,\max} &\leq 0 \label{eq:opt_8}\\
\eta_i &\geq 0 \label{eq:opt_9}\\
\eta_i(y_i p_{i,\max} - p_i) &= 0 \label{eq:opt_10}
\end{align}

Now we analyze a few different cases. Assume for now that $\lambda \neq 0$ and so we have $\sum_i p_i = L$ by \Cref{eq:opt_4}. Then for every index $j$ for which $p_j \notin \{p_{j, \min}, p_{j,\max}\}$ we have $\xi_j = \eta_j = 0$ by \Cref{eq:opt_7,eq:opt_10}, hence $(\partial C/\partial p_j) - \lambda = 0$ by \Cref{eq:opt_1}. We conclude $\partial C/\partial p_j = \partial C/\partial p_{k}$ for any $p_j, p_k$ not taking their maximum or minimum possible value. On the other hand, if $y_i \neq 0$ and $p_i = p_{i, \min}$ then \Cref{eq:opt_6,eq:opt_7} reduce to the condition $\xi_i \geq 0$ and \Cref{eq:opt_1} implies $\partial C/\partial p_i - \lambda - \xi_i = 0$. In this case, we conclude $\lambda = \partial C / \partial p_i - \xi_i$, hence $\lambda \leq \partial C / \partial p_i$. A similar argument gives that for any $i$ with $p_i = p_{i,\max}$ and $y_i \neq 0$ we have $\lambda = \partial C / \partial p_i + \eta_i$, hence $\lambda \geq \partial C / \partial p_i$. Then, setting $\lambda = \mathcal{D}$ we see that one of conditions \labelcref{itm:derivative_cond_1,itm:derivative_cond_2,itm:derivative_cond_3} from \Cref{subsec:reducing UC to knapsack} must be satisfied for every $p_i$ with $y_i = 1$.

We now consider the case $\lambda = 0$. In this case, \Cref{eq:opt_1} gives
\[
\frac{\partial C}{\partial p_i} - \xi_i + \eta_i = 0.
\]
By \Cref{eq:opt_6,eq:opt_8}, we have $\xi_i, \eta_i \geq 0$. Since $\partial C/\partial p_i \geq 0$, this equality implies that we must have $\eta_i = 0$ and $\xi_i = \partial C/\partial p_i$. Complementary slackness~\Cref{eq:opt_7,eq:opt_10} then forces $p_i = p_{i,\min}$ for all $i$. In this situation, condition~\labelcref{itm:derivative_cond_2} can be satisfied for all $p_i$ by choosing $\mathcal{D} \le \min_i \partial C/\partial p_i$.

In either case, we conclude that conditions~\labelcref{itm:derivative_cond_1,itm:derivative_cond_2,itm:derivative_cond_3} are satisfied by all $p_i$.
\section{Performance comparison across IBM Quantum processors and execution stacks}
\label{comparison}

The experiments in the main body of the paper were run on \texttt{ibm\_kingston} using Performance Management Qiskit function. In this appendix, we provide a brief comparison of the results for both the 100-qubit instance and the 150-qubit instance  on \texttt{ibm\_quebec}, without using  the Performance Management Qiskit function \cite{fire_opal, qiskit-function}. We note that we re-did the parameter training (following the procedure outlined in the main body of the paper), before running these instances on \texttt{ibm\_quebec}.
\\

In lieu of the Performance Management function, the experiments performed on \texttt{ibm\_quebec} were performed on circuits that were transpiled with the \texttt{qiskit} (v2.0.3) transpiler with \texttt{optimization\_level=3}, using the \texttt{SamplerV2} primitive from \texttt{qiskit-ibm-runtime} (v0.40.1), with $10^{5}$ shots, and with gate twirling, measurement twirling, and dynamical decoupling enabled \footnote{See, e.g, \cite{emdocs} for details on the implementation of these techniques.}. For gate twirling, we made use of the \texttt{active-circuit} twirling strategy while for dynamical decoupling, we used the \texttt{XY4} sequence. All other settings for each of these were kept at their default values. Note, in particular, that for our experiments on \texttt{ibm\_quebec} we did not make use of fractional gates (see, e.g, \cite{fr}), to allow us to use the implementation of pauli twirling that is available via \texttt{qiskit-ibm-runtime}. 
\\

Tables.~\ref{tab:circuit_props} and
~\ref{tab:performance} show a comparison of these hardware experiments. For the 100-qubit instance, each column contains a list of 5 values that correspond to cop-QAOA circuits from $p=1$ through $p=5$. 
\begin{table}[h]
\centering
\small
\begin{tabular}{|l|l|c|c|}
\hline
Processor & Instance & 2q-gate count & 2q-gate depth \\
\hline
\multirow{2}{*}{\texttt{ibm\_quebec}} 
& 100q & [190,380,570,760,950] & [4,8,12,16,20] \\
& 150q & 538 & 28 \\
\hline
\multirow{2}{*}{\texttt{ibm\_kingston}} 
& 100q & [156,159,159,158,159] & [5,7,7,5,6] \\
& 150q & 497 & 28 \\
\hline
\end{tabular}
\caption{2q-gate depth and count across two different quantum processors and execution stacks.}
\label{tab:circuit_props}
\end{table}
As is evident from this table, the Performance Management function generically provides a  reduction in the two-qubit gate counts and the two-qubit gates of the transpiled circuits. This is likely, at least in part, due to the fact that we did not use fractional gates for the experiments on \texttt{ibm\_quebec} while experiments performed using the Performance Management function on \texttt{ibm\_kingston} did make use of use fractional gates. We note that the results on both \texttt{ibm\_quebec} (with error suppression)  and \texttt{ibm\_kingston} (with performance management) are very similar. Moreover, even with less optimized transpilation that resulted in 1900 2q-gates and a 2q-gate depth of 60 for the 100q instance and 860 2q-gates and a 2q-gate depth of 53 for the 150q instance, on \texttt{ibm\_quebec}, we observed that the ratio \texttt{Best/Gurobi} was very similar to the values reported in Table.~\ref{tab:performance}. This reflects the stability and robustness of our algorithm on hardware. We also observed that increasing the number of shots to $10^{6}$ improves the hardware results even more and the algorithm shows the potential to consistently outperform Gurobi. In particular, for the 100q instance on \texttt{ibm\_quebec}, with $10^{6}$ shots, the ratio \texttt{Best/Gurobi} exceeds 1 across all values of $p$.
\begin{table*}
\centering
\small
\begin{tabular}{|l|l|c|c|}
\hline
Processor & Instance & Best/Gurobi & Valid sample ratio \\
\hline
\multirow{2}{*}{\texttt{ibm\_quebec}} 
& 100q & [0.9999790,0.9999996,0.9999790,0.9999893,0.9999894] & [0.44,0.23,0.17, 0.11,0.06] \\
& 150q & 0.97 & 0.08 \\
\hline
\multirow{2}{*}{\texttt{ibm\_kingston}} 
& 100q & [0.99997908, 0.99999968, 0.99999964, 1.00001001, 0.9999997] & [0.37,0.23,0.22,0.19,0.19] \\
& 150q & 0.97 & 0.13 \\
\hline
\end{tabular}
\caption{Performance comparison across two different quantum processors and execution stacks. Number of shots in both cases is $10^5$.}
\label{tab:performance}
\end{table*}





                    


 \end{document}